# Temperature behavior of graphene conductance induced by piezoelectric effect in ferroelectric substrate

Anna N. Morozovska[1,2*], Anatolii I. Kurchak[3], Zhanna G. Zemska[4] and Maksym V. Strikha[3,5†]

[1] *Institute of Physics, National Academy of Sciences of Ukraine,*

*Prospect Nauky 46, 03028 Kyiv, Ukraine*

[2] *Bogolyubov Institute for Theoretical Physics, National Academy of Sciences of Ukraine,*

*14-b Metrolohichna street, 03680 Kyiv, Ukraine*

[3] *V.Lashkariov Institute of Semiconductor Physics, National Academy of Sciences of Ukraine,*

*Prospect Nauky 41, 03028 Kyiv, Ukraine,*

[4] *Taras Shevchenko Kyiv National University, Physical Faculty*

*Prospect Akademika Hlushkova 4a, 03022 Kyiv, Ukraine.*

[5] *Taras Shevchenko Kyiv National University, Radiophysical Faculty*

*Prospect Akademika Hlushkova 4g, 03022 Kyiv, Ukraine.*

## Abstract

Graphene-on-ferroelectric structures can be promising candidates for advanced field effect transistors, modulators and electrical transducers, providing that research of their electro-transport and electro-mechanical performances can be lifted up from mostly empirical to prognostic theoretical level. Recently we have shown that alternating piezoelectric displacement of the ferroelectric domain surfaces can lead to the alternate stretching and separation of graphene areas at the steps between elongated and contracted domains, and the conductance of graphene channel can be increased essentially at room temperature, because electrons in the stretched section scatter on acoustic phonons. The piezo-mechanism of graphene conductance control requires systematic studies of the ambient condition impact on its manifestations. This theoretical work studies in details the temperature behavior of the graphene conductance changes induced by piezoelectric effect in a ferroelectric substrate with domain stripes. We revealed the possibility to control graphene conductance (that can change in 5 – 100 times for PZT ferroelectric substrate) by tuning the ambient temperature $T$ from low values to the critical one $T_{cr}$ for given gate voltage $U$ and channel length $L$. Also we demonstrate the possibility to control graphene conductance changes up to one hundred of times by tuning the $U$ from 0 to the critical value $U_{cr}$ at a given $T$ and $L$. The critical parameters $T_{cr}$ and $U_{cr}$ correspond to complete graphene separation above contracted domains. We derived analytical expressions for the dependence of the critical voltage $U_{cr}$ and corresponding graphene conductance on $T$, $L$ and material parameters of graphene-on-ferroelectric structure. Obtained results can open the way towards graphene-on-ferroelectric applications in piezo-resistive memories operating in a wide temperature range.

* Corresponding author. E-mail: anna.n.morozovska@gmail.com

† Corresponding author. E-mail: maksym.strikha@gmail.com

# I. INTRODUCTION

Since graphene was obtained experimentally [1, 2] till nowadays graphene and other related 2D-semiconductors [3, 4, 5, 6, 7, 8, 9] remain in focus of intensive studies, at that and the most recent works [8-9] indicate that the potential for studying the material is much higher and further fundamental and applied researches of its electro-conductive properties are highly relevant.

Most challenges associated with the practical usage of graphene, its super-lattices and other 2D-semiconductors critically depend on the optimal choice of electromechanical, electrophysical and physicochemical properties of their interfaces, substrates and gates. In particular the problem of the substrates with additional functionality degrees compatible with a given 2D-material is crucial for 2D physics nowadays [3, 10].

It should be noted that various ferroelectric substrates with a graphene channel remain promising for research, utilizing the fact that a ferroelectric significantly affects the properties of graphene [11, 12, 13, 14, 15], because spontaneous polarization and domain structure can be controlled by an external electric field [16, 17]. Note, that the polarization direction in ferroelectric substrate can be reversed by the voltage applied to the gate of the graphene field effect transistor (**GFET**), where a graphene is a channel. The $180^{o}$-ferroelectric domain walls (**FDW**) in a ferroelectric substrate induces the formation of the barriers (p-n junctions) between the regions enriched with holes and electrons in graphene, which appeared near the contact of the domain walls with the ferroelectric substrate surface [18, 19, 20]. The p-n junctions in graphene have been realized earlier by multiple gates doping of graphene channel by electrons or holes, respectively [21, 22, 23]. Then they have been studied theoretically [24, 25] and experimentally [26, 27, 28]. Due to the charge separation by an electric field of a FDW – surface junction [29, 30], p-n junction can occur without any additional gates doping and different types of carrier transport (ballistic, diffusive, etc) are possible in a graphene channel at $180^{o}$-FDWs [18, 19, 20]. At that the movement of the domain walls in the substrate it is reflected in the form of I-V hysteresis loops of GFET systems and their dynamics [18, 19, 20].

Hence graphene-on-ferroelectric structures can be promising candidates for advanced field effect transistors, modulators and electrical transducers, providing that research of their electro-transport and electro-mechanical performances can be lifted up from mostly empirical to prognostic theoretical level.

It is well-known that the elastic strain can change the band structure of graphene and open its band gap [4, 5, 31, 32, 33]. One of the insufficiently investigated effect is the change of graphene conductivity due to the partial exfoliation of graphene from the ferroelectric substrate when applying voltage to the gate of GFET, due to the presence of the piezoelectric effect in the substrate. Recently we have shown that alternating piezoelectric displacement of the ferroelectric domain surfaces can lead to the alternate stretching and separation of graphene areas at the steps between elongated and contracted domains, and the conductance of graphene channel can be increased essentially at room temperature, because electrons in the stretched section scatter on acoustic phonons [34]. We revealed a nontrivial temperature behavior of the carriers concentration, which governs the conductance of the graphene channel on ferroelectric substrate with domain walls [35]. It appeared that these effects originate from the nonlinear screening of ferroelectric polarization by graphene carriers, as well as it is conditioned by the temperature evolution of the domain structure kinetics in ferroelectric substrate. However the piezoelectric mechanism of graphene conductance control requires systematic studies of the ambient condition impact on its manifestations. This theoretical work studies in details the temperature behavior of the graphene conductance changes induced by piezoelectric effect in a ferroelectric substrate with domain stripes.

## II. THEORETICAL BACKGROUND

Performed calculations [34] have shown that one domain elongates and another one contracts depending on the voltage polarity when the voltage $U$ is applied to a gate of the GFET with FDW [compare **Fig. 1(a)** with **1(b)**]. Corresponding surface displacement $h$ can be significant and reach $(0.5 - 1)$nm for ferroelectrics with high piezoelectric coefficients like $PbZr_{0.5}Ti_{0.5}O_3$ (**PZT**). The vertical scale in plot **(b)** is $\sim (50 - 500)$ pm depending on the film thickness and temperature, while the lateral scale is typically much larger $(50 - 500)$ nm, or even microns (see e.g. **Fig.2** in Ref.[34]). The scales are so different because uncharged domain walls are usually thin $(1 - 5 \text{ nm})$ and domain period is much larger $(10 \text{ nm} - 1 \text{ } \mu\text{m})$ depending on temperature and ferroelectric film thickness. Dashed curves in plot **(a)** show the FDW broadening that appears near the surface [36].

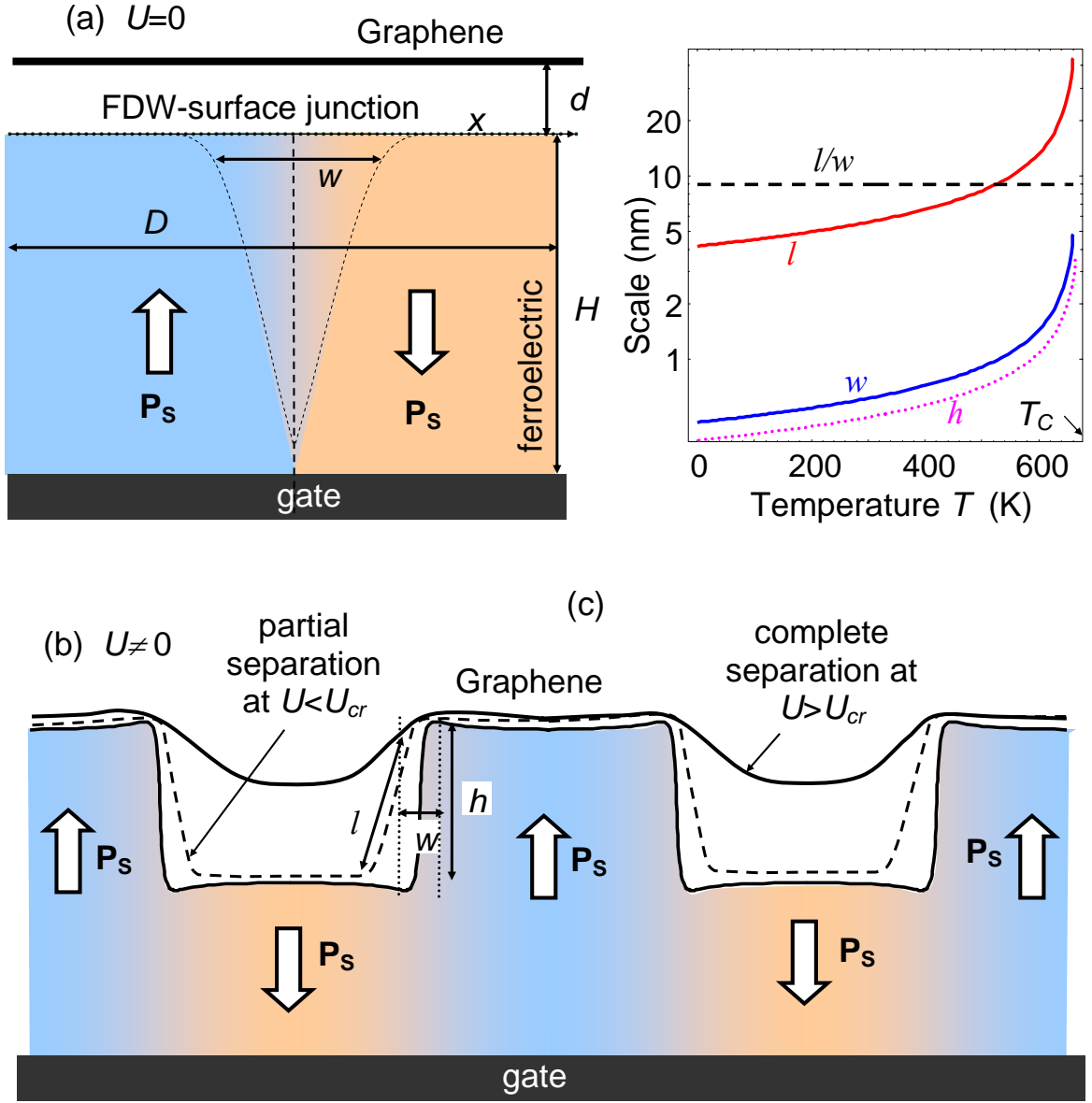

**FIG. 1.** Partial separation of graphene channel sections induced by a piezoelectric effect at the ferroelectric domain wall − surface junction. The separation is absent at $U$=0 **(a)** and appears at $U$≠0 **(b)**, where partial separation [dashed curves in plot **(b)**] or complete exfoliation [solid curves in plot **(b)**] of graphene becomes possible. For complete exfoliation the maximal length of exfoliated regions is equal to the length of domains which contract, i.e. the length $l(T,U) < L/2$. Dashed curves in plot **(a)** show the FDW broadening near the surface. **(c)** Dependence of the domain wall width $w$ (blue curve) and separated length $l$ (red curve), their ratio $l/w$ (black dashed line) and surface displacement $h$ (magenta dotted curve) on temperature for graphene-on-PZT parameters and gate voltage $U$=1V.

When the voltage $U$ is applied to a gate of the GFET with FDW, one domain elongates and another one contracts depending on the voltage polarity. The conductance of graphene channel in diffusion regime can change essentially, because electrons in the separated stretched section scatter on acoustic phonons, while more intensive scattering channel dominates in bounded sections [37]. In particular the voltage dependence of the graphene channel conductance $G(T,U)$ corresponding to the case when its part of length $l(U)$ is separated and another part of length $L - l(U)$ is bounded, obeys the Matiessen rule [37]:

$$G(T,U) = \begin{cases} W\left(\dfrac{L - l(T,U)}{\sigma_B(T)} + \dfrac{l(T,U)}{\sigma_S(T)}\right)^{-1}, & L \geq 2l(T,U) \\ W\dfrac{L}{2}\left[\dfrac{1}{\sigma_B(T)} + \dfrac{1}{\sigma_S(T)}\right]^{-1}, & L < 2l(T,U) \end{cases} \quad (1)$$

Here $L$ is graphene channel length and $W$ is its width. The expression for separated length $l$ was derived in Ref [34]. Here we derived its temperature dependence in the form

$$l(T,U) = \sqrt[3]{\dfrac{4Yd}{J}}\left[d_{33}(T) + (1+2v)d_{31}(T)\right] \cdot |U| \approx \begin{cases} \sqrt[3]{\dfrac{4Yd}{J}}\dfrac{\left(d_{33}^0 + (1+2v)d_{31}^0\right)}{\sqrt{1 - T/T_C}}|U|, & T < T_C \\ 0, & T > T_C \end{cases}, \quad (2)$$

Here $d$ is the distance between the flat surface of ferroelectric and graphene, and $v$ is the Poisson ratio [38]. The thickness $d$ of the physical gap between the graphene and ferroelectric is determined by Van-der-Waals interaction, and its value is less than 1 nm. The density $J$ of binding energy for graphene on ferroelectric should be smaller that the one for $SiO_2$ substrate because graphene adhesion to mica surface is considered to be the strongest one in comparison with other surfaces (0.5 J/m$^2$) [39]. On the contrary the Young's modulus of graphene is extremely high ($Y = 1$ TPa [40, 41]). Constants $d_{33}$ and $d_{31}$ are temperature-dependent piezoelectric coefficients, $d_{33}(T) = \dfrac{d_{33}^0}{\sqrt{1 - T/T_C}}$ and $d_{31}(T) = \dfrac{d_{31}^0}{\sqrt{1 - T/T_C}}$, which increase when approaching Curie temperature and then becomes zero at $T > T_C$ [42]. For PZT $d_{33}^0 = 741.316$ pm/V, $d_{31}^0 = -333.592$ pm/V, $T_C = 666$ K [43]. Note that the linear approximation Eq.(2) of the separated graphene region, $l(T,U) \sim |U|$, can be used for the case when the length $l$ is at least longer than the lateral halfwidth $w/2$ of the ferroelectric surface displacement step at the FDW [see **Fig.1(b)**]. Since the width $w$ is restricted by the intrinsic width of the uncharged domain wall that's temperature dependence is given by expression $\sqrt{g/\alpha_T(T_C - T)} \sim w_0/\sqrt{1 - T/T_C}$, where $w_0 = \sqrt{g/\alpha_T T_C}$ is the intrinsic width of uncharged domain wall at 0 K that is about lattice constant (~0.5 nm for PZT). So that the ratio $l(T,U)/w$ is temperature independent and equal to $\sqrt[3]{\dfrac{4Yd}{J}}\dfrac{|U|}{w_0}\left(d_{33}^0 + (1+2v)d_{31}^0\right)$ in accordance with Eq.(2) [see **Fig.1(c)**].

The critical temperature corresponding to the condition $2l(T_{cr}, U) = L$ is given by expression

$$T_{cr}(L,U) = T_C \left[ 1 - \left( \frac{2|U|}{L} \cdot \sqrt[3]{\frac{4Yd}{J}} \left( d_{33}^0 + (1+2\nu) d_{31}^0 \right) \right)^2 \right] \qquad (3a)$$

At fixed temperature $T < T_C$ the critical voltage corresponding to the condition $2l(T, U_{cr}) = L$ is given by expression

$$|U_{cr}| = \sqrt[3]{\frac{J}{4Yd}} \frac{\sqrt{1 - T/T_C}}{d_{33}^0 + (1+2\nu) d_{31}^0} \frac{L}{2}, \qquad T < T_C. \qquad (3b)$$

The critical temperature and gate voltage correspond to complete separation of graphene above the contracted domains [see solid curves in **Fig.1(b)**].

## III. ANALYSIS OF THE CRITICAL GATE VOLTAGE AND TEMPERATURE

**Figure 2(a)** is the color map of the critical voltage $U_{cr}(T, L)$ plotted in coordinates "temperature T − channel length L". The value of $U_{cr}$ varies from 0 at $T = T_C$ to 60 V at $T = 0$ K and $L$=500 nm. At fixed $L$ the value of $U_{cr}$ monotonically increases with the temperature decrease, because $|U_{cr}| \sim \sqrt{1 - T/T_C}$ in accordance with Eq.(3b). At fixed $T$ the value of $U_{cr}$ linearly increases with L increase, because $|U_{cr}| \sim L$ in accordance with Eq.(3b). **Fig. 2(a)** demonstrates the possibility to control graphene maximal separation by tuning the gate voltage $U$ from 0 to $U_{cr}$ for given temperature $T$ and channel length $L$.

**Figure 2(b)** is the color map of the critical temperature $T_{cr}(U, L)$ plotted in coordinates "gate voltage $U$ − channel length $L$". The value of $T_{cr}$ varies from 207 K at $U = \pm 10$ V and $L$=50 nm to $T_C$=666 K at $U = 0$ and L=500 nm. At fixed $L$ the value of $T_{cr}$ monotonically decreases with the gate voltage increase, because $T_{cr}(L,U) = T_C \left[ 1 - \left( \frac{2|U|}{L} \cdot C \right)^2 \right]$ in accordance with Eq.(3a), $C$ is a material constant. At fixed $U$ the value of $T_{cr}$ increases with L increase as $1/L^2$ in accordance with Eq.(3a). **Fig. 2(b)** demonstrates the possibility to control graphene maximal separation by tuning the ambient temperature $T$ from 200 K to Curie temperature for given gate voltage $U$ and channel length $L$.

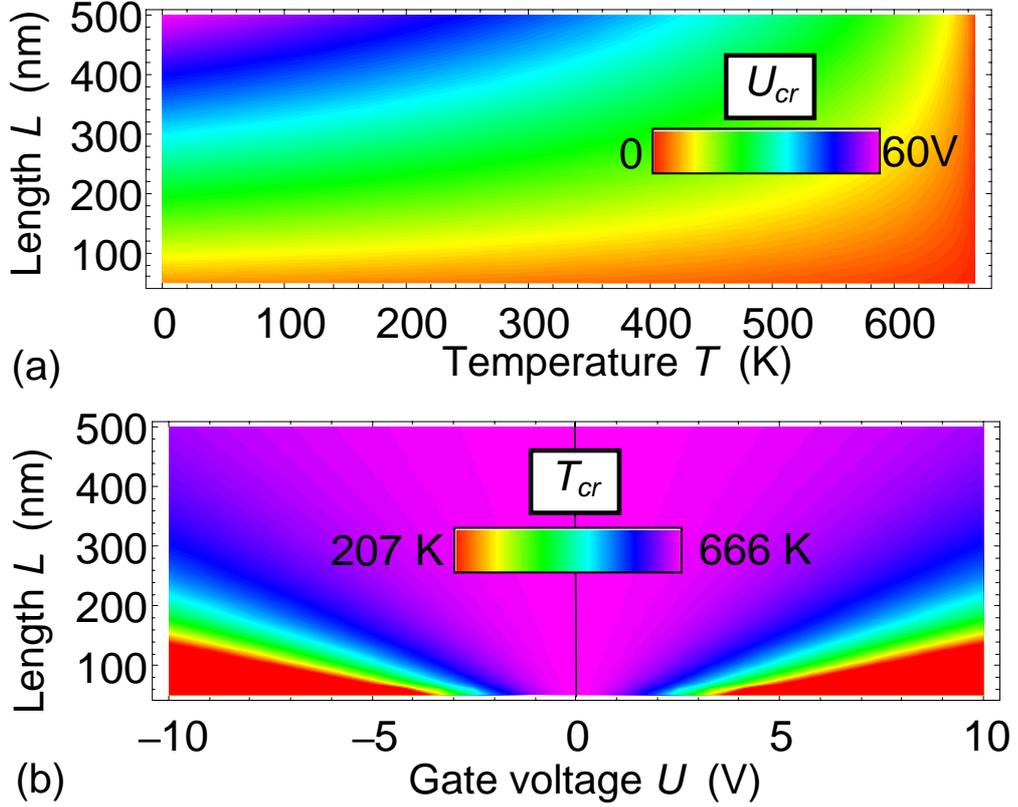

**FIG. 2. (a)** Color map of the critical voltage $U_{cr}(T, L)$ plotted in coordinates "temperature $T$ – channel length $L$". **(b)** Color map of the critical temperature $T_{cr}(U, L)$ plotted in coordinates "gate voltage $U$ – channel length $L$". Piezoelectric coefficients $d_{33}^0$ =741.316 pm/V, $d_{31}^0$ =-333.592 pm/V, Curie temperature $T_C$ =666 K and Poisson ratio $\nu$=0.3 corresponds to PbZr$_{0.5}$Ti$_{0.5}$O$_3$. Graphene-ferroelectric and separation $d = 0.5$ nm, binding energy $J$=0.25 J/m$^2$, graphene Young's modulus $Y = 1$ TPa. Film thickness $H$>100 nm. Color scale ranges from minimal (red color) to maximal (violet color) values indicated by numbers.

## IV. TEMPERATURE AND LENGTH BEHAVIOUR OF GRAPHENE CHANNEL CONDUCTANCE

The conductivity of the bounded section, where the scattering of electrons at substrate ionized impurities dominate, has the form listed in Ref.[34]. Corresponding temperature dependence has the form:

$$\sigma_B(T) = \frac{2e^2}{\pi^{3/2}\hbar}\lambda_B[n_S]\sqrt{n_S} = \frac{2e^2\alpha}{\pi^{3/2}\hbar}n_S \approx \frac{2e\alpha}{\pi^{3/2}\hbar}P_S^0\sqrt{1-\frac{T}{T_C}}, \qquad T < T_C. \qquad (4)$$

Here $e$=1.6×10$^{-19}$ C is elementary charge, $\hbar = 1.056\times10^{-34}$ J·s = $6.583\times10^{-16}$ eV·s is Plank constant, $\nu_F = 10^6$ m/s is characteristic electron velocity in graphene, $\lambda_B[n_S] = \alpha\sqrt{n_S}$ is mean free path in the graphene channel, and the proportionality coefficient α depends on the substrate material and graphene-ferroelectric interface chemistry. The concentration of 2D electrons $n_S$ can be regarded constant voltage-independent value far from the FDWs and proportional to the spontaneous polarization, $P_S(T)$, namely $n_S \approx |P_S/e|$ and $P_S(T) = P_S^0 \sqrt{1 - \dfrac{T}{T_C}}$.

Conductivity of the separated stretched section of structurally perfect graphene is ruled by collisions with acoustic phonons $\sigma_S$ and doesn't depend on 2D electrons concentration in the graphene channel. (see e.g. Refs.[21] and [34]). The upper limit for $\sigma_S$ is:

$$\sigma_S(T) = \frac{4e^2 \hbar \rho_m v_F^2 v_S^2}{\pi D_A^2 k_B T} \qquad (5)$$

In Eq.(5) $\rho_m \approx 7.6 \cdot 10^{-7}\,\text{kg/m}^2$ is 2D mass density of carriers in graphene, $v_S \approx 2.1 \cdot 10^4\,\text{m/s}$ is a sound velocity in graphene, Boltzmann constant $k_B = 1.38 \times 10^{-23}\,\text{J/K}$, $D_A \approx 19\,\text{eV}$ is acoustic deformation potential that describes electron-phonon interaction [34].

**Figures 3(a-d)** illustrate the dependences of the conductance $G(U,T)$ on the gate voltage $U$ calculated for several temperatures $T$ in the range (100- 655) K and channel length $L$ in the range $(50 - 500)$ nm**.** From the figures the conductance, being a monotonically increasing function of $|U|$ in accordance with Eqs.(1)-(2), becomes voltage independent at $|U| > |U_{cr}|$. At that the U-type curves of $G(U,T)$ are the highest for the lowest temperature 100 K and monotonically sink down with the temperature increase to 655 K (compare the curves $1 - 7$). The conductance decrease is more than 5 times with the temperature change from 100 K to 655 K. In accordance with Eq.(3b) and **Fig.2(a)** the value of $|U_{cr}|$ increases with $L$ increase from 1 V to 10 V and higher. Because of the fact the voltage region of U-type curves of $G(U,T)$ significantly increases with $L$ increase from 50 nm to 500 nm [compare plots $(a) - (d)$]. **Figs. 3** demonstrates a promising role of piezoelectric effect to control the voltage dependence of graphene channel conductance by tuning the gate voltage $U$ from 0 to $U_{cr}$ for given temperature $T$ and channel length $L$.

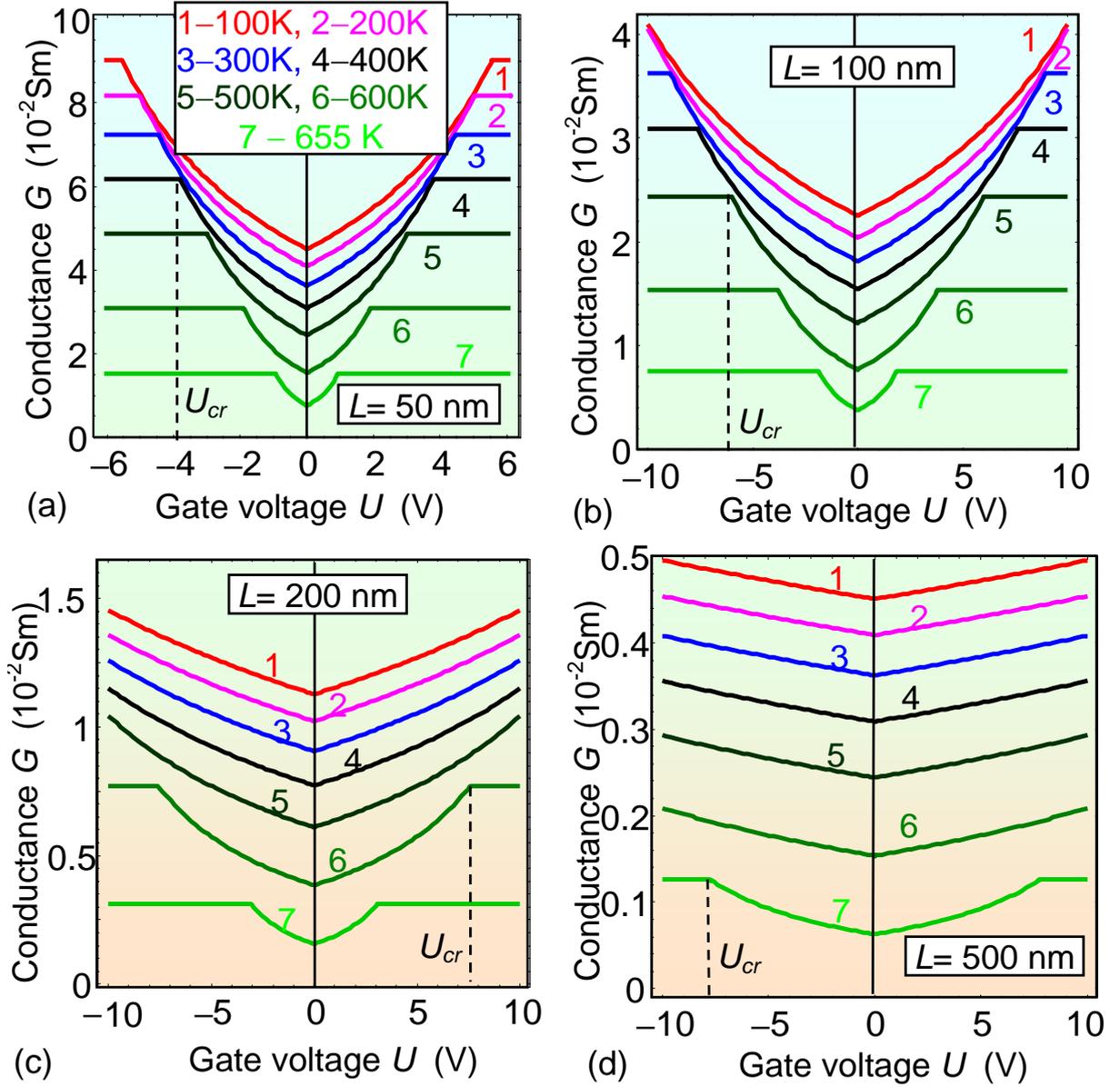

**FIG. 3.** Dependences of the conductance $G(U,T)$ on the gate voltage $U$ calculated for several temperatures $T$=100, 200, 300, 400, 500, 600 and 655 K (curves 1 − 7) and channel length $L$ = 50 nm, **(a)** 100 nm **(b)**, 200 nm **(c)**, 500 nm **(d)**. Electron mean free path $\lambda_B$ =10 nm and concentration $n_S$ = 3×10$^{18}$ m$^{-2}$ at $T = 0$, channel width $W$= 50 nm. Other parameters are the same as in **Fig. 2**.

**Figure 4(a)** is the color map of the graphene channel conductance $G(T,U)$ plotted in coordinates "gate voltage $U$ − channel length $L$" calculated at room temperature. The value of $G(T,U)$ varies from 3.6 mSm for $U$=0 and $L$=500 nm to 72.7 mSm for $|U|$ =6V and $L$=50 nm. At fixed $L$ the value of $G(T,U)$ relatively slowly and quasi-linearly increases with the $U$ increase until $|U| < |U_{cr}(L,T)|$, where $|U_{cr}(L,T)|$ follows from Eq.(3b). At fixed gate voltage $|U| < |U_{cr}(L,T)|$ the value of $G(T, U)$ significantly decreases with $L$ increase, because the ratio $l/L$ decreases in accordance with Eq.(1).

**Figure 4(b)** is color map of the conductance $G(T, U)$ plotted in coordinates "gate voltage $U$ – channel length $L$" calculated at temperature 660 K that is very close to $T_C$. The value of $G(T, U)$ varies from 0.5 mSm for $U=0$ and $L=500$ nm to 9.3 mSm for $|U| = 6$V and $L=50$ nm. At fixed $L$ the value of $G(T, U)$ noticeably quasi-linearly increases with the $U$ increase until $|U| < |U_{cr}(L, T)|$, where $|U_{cr}(L, T)|$ follows from Eq.(3b). At fixed gate voltage $|U| < |U_{cr}(L, T)|$ the value of $G(T, U)$ significantly decreases with $L$ increase, because the ratio $l/L$ decreases in accordance with Eq.(1). Despite the color map 4(b) demonstrates significant changes with $U$ increase, but the absolute values of the conductance at 660 K are much lower than the ones at room temperature.

Hence **Figs. 4** demonstrate the possibility to control graphene channel conductance by tuning the gate voltage $U$ and channel length $L$ for a given temperature $T$.

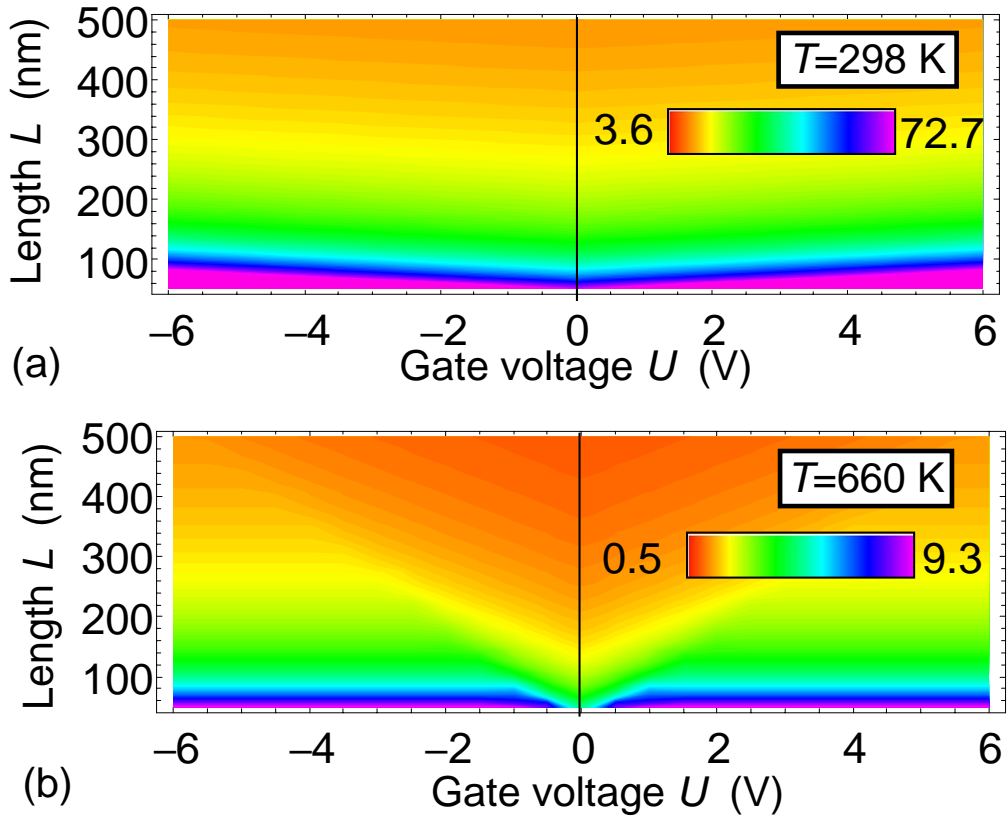

**FIG. 4.** Color maps of the conductance $G(T, U)$ (in mSm) plotted in coordinates "gate voltage U – channel length $L$" calculated at room temperature 298 K **(a)** and 660 K **(b)**. Other parameters are the same as in **Fig. 2.** Color scale ranges from minimal (red color) to maximal (violet color) values indicated by numbers.

**Figure 5(a)** is the color map of the graphene channel conductance $G(T,U)$ plotted in coordinates "gate voltage $U$ – temperature $T$" calculated for $L$=50 nm. The value of $G(T,U)$ varies from 0 for $T=T_C$ to 97.8 mSm for $|U|=10$V and $L$=50 nm. At fixed $T$ the value of $G(T,U)$ quasi-parabolically increases with the $U$ increase until $|U|<|U_{cr}(L,T)|$, where $|U_{cr}(L,T)|$ follows from Eq.(3b). At fixed gate voltage $|U|<|U_{cr}(L,T)|$ the value of $G(T,U)$ significantly decreases with $T$ increase, because the ratio $l/L$ decreases in accordance with Eq.(1).

**Figure 5(b)** is color map of the conductance $G(T,U)$ plotted in coordinates "gate voltage $U$ – temperature $T$" calculated for $L$=100 nm. The value of $G(T,U)$ varies from 0 for $T=T_C$ to 41.8 mSm for $|U|=10$V and $L$=100 nm. At fixed $T$ the value of $G(T,U)$ noticeably increases with the $U$ increase until $|U|<|U_{cr}(L,T)|$.

**Figure 5(c)** is color map of the conductance $G(T,U)$ plotted in coordinates "gate voltage U "gate voltage $U$ – temperature $T$" calculated for $L$=500 nm.. The value of $G(T,U)$ varies from 0 for $T=T_C$ to 5.3 mSm for $|U|=10$V and $L$=500 nm. At fixed $T$ the value of $G(T,U)$ slightly quasi-linearly increases with the $U$ increase, the critical voltage is high and not shown in the figure.

Hence **Figs. 5** demonstrate the possibility to control graphene channel conductance by tuning the gate voltage U and temperature $T$ for a given channel length L.

To resume numerical results performed for graphene on PZT ferroelectric substrate and shown in **Figs.2-5**, we note that graphene channel conductance can be changed in 5-100 times by changing the temperature from low values to either the critical value $T_{cr}$ or to ferroelectric Curie temperature (666 K for PZT) depending on the gate voltage $U$ and channel length $L$. Also we demonstrate the possibility to control graphene conductance changes up to 100 times by tuning the $U$ from 0 to the critical value $U_{cr} \sim (1-10)$ V at a given $T$ and $L$.

All above numerical results were performed for a thick PZT substrate with relatively high $T_C$ = 666 K. Notably the Curie temperature $T_{cr}(H)$ of the substrate can be tuned by changing its thickness $H$ due to the finite size effects [44], $T_{cr}(H) \approx T_C - \dfrac{d}{\alpha_T \varepsilon_0 \varepsilon_d H}$ [20, 30, 34]. From the expression it can be close to the room temperature. Another ferroelectric material can be chosen for substrate with e.g. lower Curie temperature, such as BaTiO$_3$ with $T_C$ = 381 K.

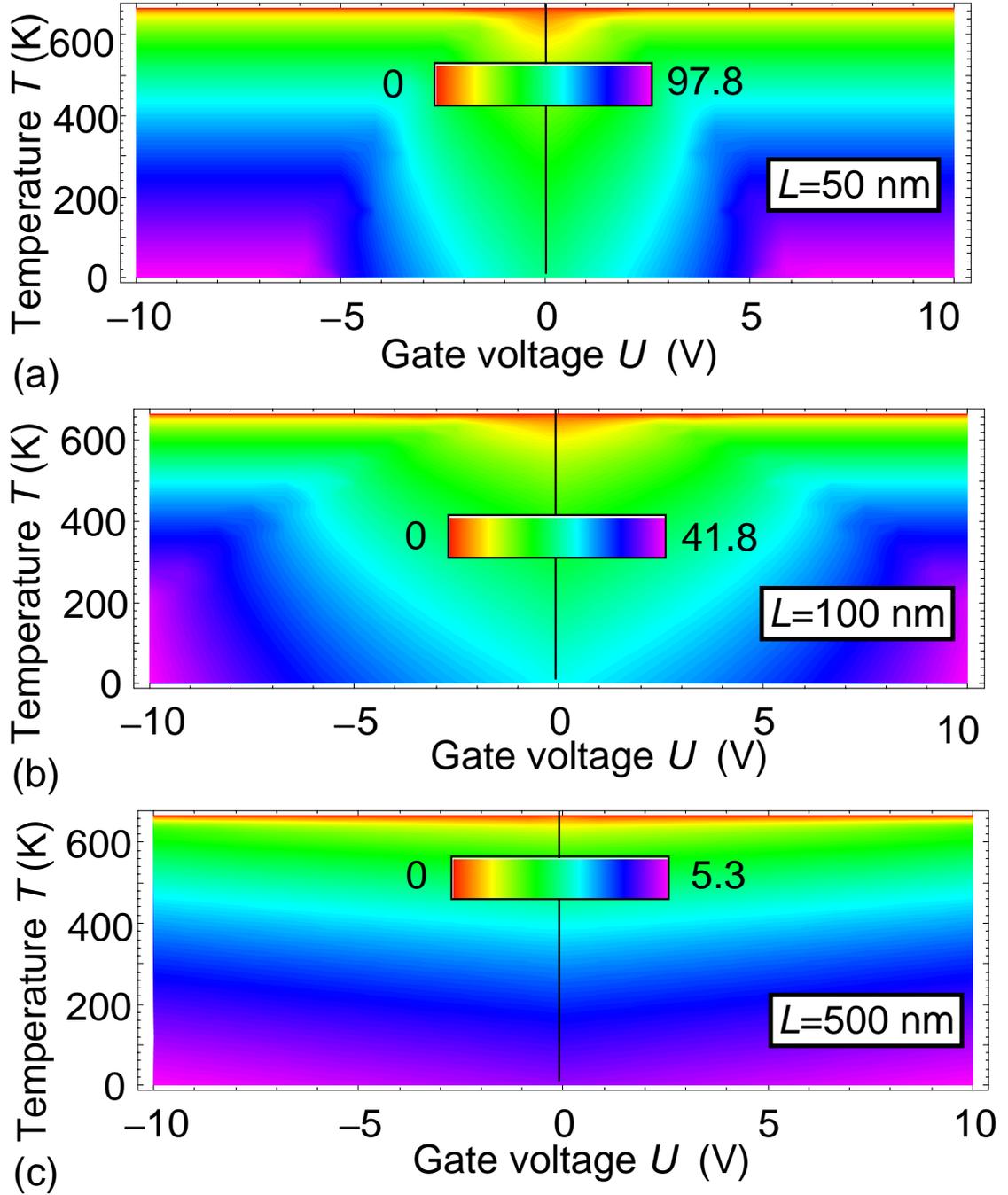

**FIG. 5.** Color maps of the conductance $G(T,U)$ (in mSm) plotted in coordinates "gate voltage U – temperature T" calculated at several channel length $L$= 50 nm **(a)** 100 nm **(b)** and 500 nm **(c)**. Other parameters are the same as in **Fig. 2.** Color scale ranges from minimal (red color) to maximal (violet color) values indicated by numbers.

### V. SUMMARY

To summarize, graphene-on-ferroelectric structures can be promising candidates for advanced field effect transistors, modulators and electrical transducers, providing that research of their electro-transport and electro-mechanical performances can be lifted up from mostly empirical to prognostic theoretical level.

Recently revealed piezo-mechanism of graphene conductance control [34] requires systematic studies of the ambient condition impact on its manifestations. Therefore the studies of the temperature behavior of the changes of graphene conductance induced by piezoelectric effect in ferroelectric substrate with domain walls are in order.

This work demonstrate the possibility to control graphene conductance (that can change in 5 – 100 times for PZT ferroelectric substrate with domain stripes) by tuning the ambient temperature $T$ from low values to the critical one $T_{cr}$ (that can reach ferroelectric Curie temperature) for given gate voltage $U$ and channel length $L$. For PZT substrate with evolved domain structure we demonstrate the possibility to control graphene conductance changes up to 100 times by tuning the $U$ from 0 to the critical value $U_{cr}$ at a given $T$ and $L$. The critical parameters $T_{cr}$ and $U_{cr}$ correspond to complete graphene separation above contracted domains. We derived analytical expressions for the dependence of the critical voltage $U_{cr}$ and corresponding graphene conductance $G(U)$ on $T$, $L$ and material parameters of graphene-on-ferroelectric structure.

Obtained results can open the way towards graphene-on-ferroelectric applications in piezo-resistive memories operating in a wide temperature range.

**Acknowledgments.** A.N.M. work was partially supported by the National Academy of Sciences of Ukraine (projects No. 0117U002612, No. 0118U003375) and by the Program of Fundamental Research of the Department of Physics and Astronomy of the National Academy of Sciences of Ukraine (project No. 0117U000240).